\documentclass[structabstract]{aa} 
\usepackage{subfigure}                                   
\usepackage{epsfig}
\usepackage{wasysym}
\usepackage{graphicx}
\usepackage{natbib}
\usepackage{lscape}
\usepackage{multirow}

\usepackage{color}
\providecommand{\tabularnewline}{\\}
\usepackage{txfonts}
\begin{document}

\title{Orbital and physical properties of planets and their hosts: \\new insights on planet formation and evolution.}

\subtitle{}

\titlerunning{Period mass metallicity}


\author{V.~Zh.~Adibekyan\inst{1} 
\and P.~ Figueira\inst{1}
\and N.~C.~Santos\inst{1,2}
\and A.~Mortier\inst{1,2}
\and C.~Mordasini\inst{3}
\and \\E.~Delgado~Mena\inst{1}
\and S.~G.~Sousa\inst{1,4}
\and A.~C.~M.~Correia\inst{5,6}
\and G.~Israelian\inst{4,7}
\and M.~Oshagh\inst{1,2}
}

\institute{Centro de Astrof\'{\i}ísica da Universidade do Porto, Rua das Estrelas,
4150-762 Porto, Portugal\\
\email{Vardan.Adibekyan@astro.up.pt}
\and Departamento de F\'{\i}ísica e Astronomia, Faculdade de Ci\^{e}ncias da Universidade do Porto, Portugal
\and Max-Planck-Institut f\"ur Astronomie, K\"onigstuhl 17, D-69117 Heidelberg, Germany
\and Instituto de Astrof\'{\i}sica de Canarias, 38200 La Laguna, Tenerife, Spain
\and Departamento de F\'{\i}ísica, I3N, Universidade de Aveiro, Campus de Santiago, 3810-193 Aveiro, Portugal
\and ASD, IMCCE-CNRS UMR8028, Observatoire de Paris, UPMC, 77 Avenue Denfert-Rochereau, F-75014 Paris, France
\and Departamento de Astrof{\'\i}sica, Universidad de La Laguna, 38206 La Laguna, Tenerife, Spain
}

   \date{Received ... / Accepted ...}

 
  \abstract
   {}
{We explore the relations between physical and orbital properties of planets and properties of their host stars to identify the main observable
signatures of the formation and evolution processes of planetary systems.}
{We used a large sample of FGK dwarf planet-hosting stars with stellar parameters derived in a homogeneous way from the 
SWEET-Cat database to study the relation between stellar metallicity and position of planets in the period-mass diagram. We then
used all the radial-velocity-detected planets orbiting FGK stars to explore the role of planet-disk and planet-planet interaction on the evolution of orbital properties of 
planets with masses above 1$M_{Jup}$.}
{Using a large sample of FGK dwarf hosts we show that planets orbiting metal-poor stars have longer periods than those in metal-rich systems.
This trend is valid for masses at least from $\approx$10M$_{\oplus}$ to $\approx$4$M_{Jup}$.
Earth-like planets orbiting metal-rich stars always show shorter periods (fewer than 20 days) than those orbiting metal-poor stars. 
However, in the short-period regime there are a similar number of planets orbiting metal-poor stars.
We also found statistically significant evidence that very high mass giants (with a mass higher than 4$M_{Jup}$) have on average more eccentric orbits than giant
planets with lower mass. Finally, we show that the eccentricity of planets with masses higher than 4$M_{Jup}$ tends to be lower for planets with shorter periods. }
{Our results suggest that the planets in the P-M$_{P}$ diagram are evolving differently because of a mechanism that operates over a 
wide range of planetary masses. This mechanism is stronger or weaker depending on the metallicity of the respective system.
One possibility is that planets in metal-poor disks form farther out from their central star and/or they form later and
do not have time to migrate as far as the planets in metal-rich systems. 
The trends and dependencies obtained for very high mass planetary systems  
suggest that planet-disk interaction is a very important and orbit-shaping mechanism for planets in the high-mass domain.    
}


\maketitle
%

\section{Introduction}

The history of the discovery of extrasolar planets is a story of challenges for theories of planet formation. 
The first extrasolar planet orbiting a main sequence star\footnote{The first detection of two terrestrial-mass 
exoplanets around a pulsar PSR B1257+12 was announced in 1992 \citep{Wolszczan-92}.}%
, 51 Peg b \citep{Mayor-95},  proved to be very much at odds with the theory of formation of 
our own solar system (SS), our only reference at the time. As the number of planets increased and 
several close-in (hot-) Jupiters were found, it became clear that a new formation mechanism was necessary. 
An in situ formation of these planets was unlikely, because of the insufficient disk mass close to the star. 
However, their presence at an orbit of $\sim$0.05 AU could be reconciled with a farther out formation by invoking migration 
during or after the formation process \citep[e.g.][]{Lin-96}. The SS formation theories suggested that giant planets preferentially formed close to 
the ice-line (especially in low-metallicity disks), where water is condensed into ice and the necessary building blocks for the formation of planets
could be found in large quantities \citep[e.g.][]{Ida-08}. 
An important breakthrough was thus to explain the existence of hot-Jupiters by considering core-accretion 
simultaneously with disk-driven migration \citep[e.g.][]{Pollack-96, Alibert-05, Mordasini-09a} 
while reproducing several observational trends \citep{Mordasini-09b}.
It explained the correlation of the presence of giant planets with stellar metallicity 
\citep[e.g.][]{Santos-01, Santos-04, Fischer-05, Sousa-11}, a correlation that could not be explained by the 
alternative formation method, gravitational instability \citep[e.g.][]{Boss-98}. 

However, when the formation paradigm seemed to be complete, an unexpected result emerged. The discovery of 
hot-Jupiters whose orbital plane was misaligned with the stellar rotation axis 
\citep[e.g.][]{Hebrard-08, Triaud-10, Brown-12} cast serious doubts on disk-driven migration as the mechanism responsible for 
the hot-Jupiters. In particular, the occurrence of retrograde planets required an additional mechanism. Two main solutions put 
forward were Kozai cycles plus tidal perturbations \citep[e.g.][]{Wu-03,Fabrycky-07,Correia-11} and planet-planet scattering 
\citep[e.g.][]{Rasio-96,Beauge-12}. The implication was that planets first form in the disk at several AU,
then undergo gravitational perturbations that increase the eccentricity to very high values, and finally reduce their semi-major axis by tidal interactions 
with the star that simultaneously dampen the eccentricity to zero by conservation of the orbital angular momentum\footnote{
However, there are several other mechanisms that can tilt a star relative to its protoplanetary disk: gravitational torques from massive distant bodies 
\citep{Batygin-12} or angular momentum transport within the host star \citep{Rogers-12}.}.%

To study the main mechanism responsible for the presence of close-in Jupiters, \cite{Socrates-12} and \cite{Dawson-12} analyzed the eccentricity distribution of 
proto-hot Jupiters. The paucity of super-eccentric proto-hot Jupiters observed in the Kepler sample allowed the latter authors to conclude that disk migration is 
the dominant mechanism that produces hot-Jupiters, although some of these planets might be perturbed to high-eccentricity orbits by interactions with planetary companions.

It is interesting to note how the signatures of the formation and evolution mechanisms relate to the mass of 
the planet. There is an ongoing debate on whether the core-accretion model can reproduce the properties of Earth-mass 
planets \citep[][]{Fortier-13}. Recent works suggested that unlike for more massive planets, these systems might have been formed in situ 
\citep[e.g.][]{Hansen-12, Chiang-13}. While these types of studies are still in their infancy, they already reproduce 
some of the most common features of the low-mass planet systems, such as being dynamically 
packed and showing low inclination and eccentricities \citep{Lovis-11, Figueira-12}. On the other hand, planets 
with masses between 0.05$M_{Jup}$ and 20$M_{Jup}$ have been characterized by core-accretion and migration formation, 
and evolution models and can be compared quite well with the observations.

Evidence for the migration processes involved in the formation of short-period planets has previously been discussed in the literature. 
For instance, several works suggested discontinuities in observables inside this mass range. 
\cite{Beauge-13} presented tentative evidence that the smallest ($<4R_{\oplus}$) Kepler planetary candidates 
orbiting metal-poor stars show a period dependence (see Sect. 2); \cite{Dawson-13} argued that giant planets orbiting 
metal-rich stars show signatures of planet interaction. The first result provides compelling evidence for 
the importance of migration and accretion, the second for planet-planet interaction. For the upper end of the mass population, 
composed of planets with masses above 4$M_{Jup}$, a long  discussion has been ongoing on whether 
this population resembles more closely lower-mass planets or shares properties with stellar binaries 
\citep[e.g.][]{Udry-02, Halbwachs-05}. The eccentricity distribution of this planetary population has 
been shown to be similar to that of binaries, and some doubts were cast on core-accretion as their 
formation mechanism \citep[][]{Ribas-07}.


In this work we explore the relations and correlations between planetary mass and orbital parameters, 
namely period and eccentricity, and how they relate to metallicity, for a wide range of planetary masses. 
We start by extending the work of \cite{Beauge-13} for higher planetary masses in Sect. 2, and in Sect. 3 
we explore the properties of planets with masses higher than 4$M_{Jup}$. We summarize our main results and their implications 
on our understanding of planet formation in Sect. 4. 

We note that in this work the masses of the planets (M$_{P}$) are indeed minimum masses, so that in some cases the ``true`` planet mass
may be significantly higher. Nevertheless, statistical analyses show that the distribution of M$_{P}$ sin \textit{i} values is similar to that of M$_{P}$
values (e.g. \cite{Jorissen-01}, but see the discussion by \cite{Lopez-12} for low-mass planets). 
In fact, the average factor of overestimation is only 1.27, assuming a random distribution of the inclination.

We would like to say a word of caution regarding the significance estimates quoted in the next sections. 
In general, if one uses the same data to test several hypotheses, 
the  results can be affected by the multiple-testing problem\footnote{See 
\texttt{ http://en.wikipedia.org/wiki/Multiple\_comparisons}}
. However, since throughout our study we dealt with only a few parameters and the significance 
levels of the results are very high, it can be shown by simple algebra that the current proposed trends are significant, 
although we acknowledge the possibility that the quoted  significance levels might be optimistic.

\section{Period-mass diagram and metallicity}

Recently, \cite{Beauge-13} analyzed the distribution of planets in the orbital period (P) versus planetary radius (R) diagram  and P versus  M$_{P}$ diagram.
They found a lack of small-size/low-mass planets (R $\lesssim$ 4$R_{\oplus}$, M$_{P}$ $\lesssim$ 0.05$M_{Jup}$) with periods  P $<$ 5 days around metal-poor 
stars. They also found a paucity of sub-Jovian size/mass planets around metal-poor stars with periods shorter than  100 days. 
They explained these observed trends with a delayed formation and less planetary migration in metal-poor disks.
Interestingly, the authors found no significant correlation between metallicity and the position of giant planets in the (P,R) or (P,M$_{P}$) diagram. 
In this section we extend the analysis of these authors to higher-masses using a large sample of FGK dwarf stars ($M_{*} > 0.5 M_{\odot}$)
 with stellar parameters derived in a homogeneous way \citep[SWEET-Cat: a catalog of stellar parameters for stars with planets:][]{Santos-13}\footnote{
\texttt{https://www.astro.up.pt/resources/sweet-cat}}. 

\begin{figure*}
\begin{center}
\begin{tabular}{cc}
\includegraphics[angle=0,width=0.5\linewidth]{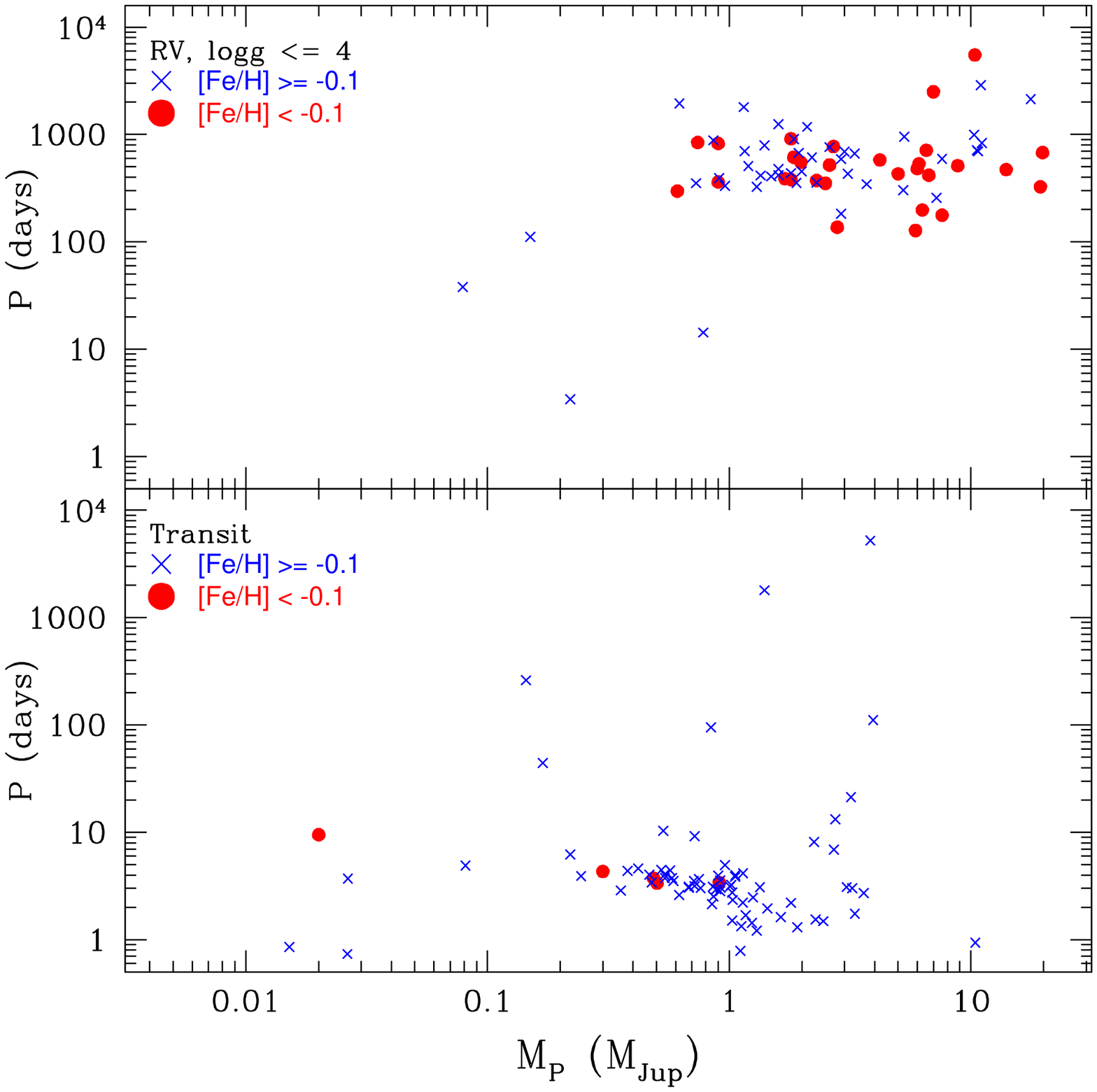} 
\includegraphics[angle=0,width=0.5\linewidth]{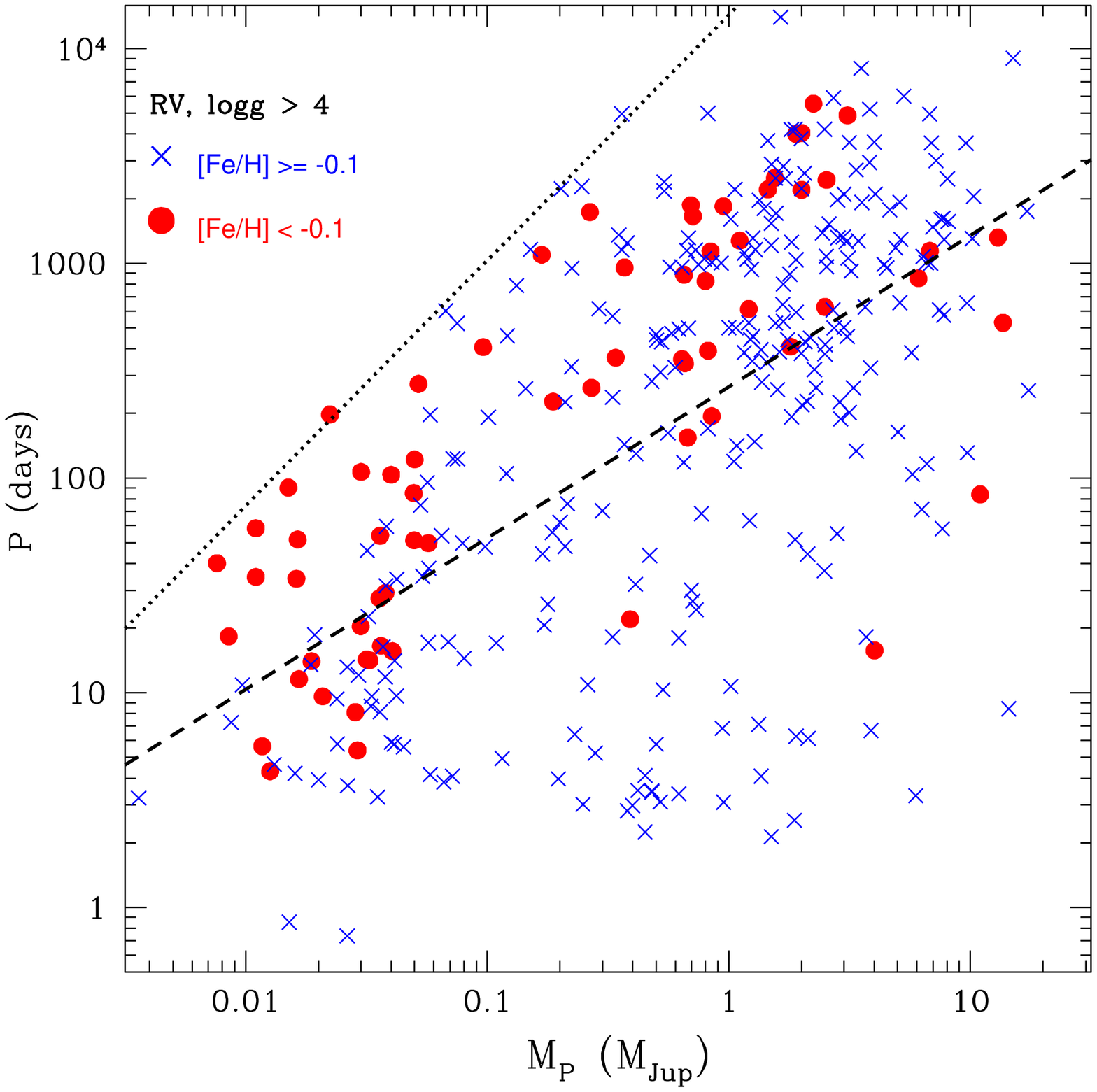}
\end{tabular}
\end{center}
\caption{Planet distribution in the (P, M$_{P}$) diagram. In the top-left panel only the planets around evolved stars with logg $\leq$ 4 dex are shown.
In the bottom-left panel we plot only the planets detected with the transiting method, and in the right panel we present the planets orbiting FGK dwarfs (logg $>$ 4 dex). 
The dotted line represents the approximate (empirical) detection limit for planets, and the dashed line is the linear fit for the full sample.}
\label{fig_p_m_feh}
\end{figure*}

Figure~\ref{fig_p_m_feh} shows the (P,M$_{P}$) diagram for planets detected with different techniques that orbit stars with different evolutionary states.
In the plot we separate the planets according to the metallicity of their host stars.
As one can see in the $bottom-left$ panel of the figure, most of the transiting planets have very short periods and their hosts are confined to a narrow range of 
metallicity (their hosts are mostly metal-rich). It is also very clear that planets around evolved stars have a very small range of P and their hosts
show very different metallicity distributions compared with their unevolved counterparts (for example, about 40\% of planets around evolved stars 
have [Fe/H] $<$ -0.1 dex, while fewer than 20\% of planets orbiting dwarf stars have similarly low metallicity). 
This difference in metallicity distribution is probably due to selection biases in evolved  stellar samples that are used for planets searches \citep{Mortier-13}.
We conclude therefore that these samples have such strong selection and detection biases that they are not suitable for our analysis.
We therefore concentrate the discussion on dwarf stars with planets detected by radial velocity when we discuss the impact of the planet host's metallicity.

The right panel of Fig.~\ref{fig_p_m_feh} shows RV-detected planets orbiting FGK dwarfs on the P-M$_{P}$ plane in which we separate low- and high-metallicity hosts
at a threshold of -0.1 dex, which is the average metallicity of the stars in the solar neighborhood \citep[e.g.][]{Adibekyan-12c}.
The same diagram, but for planets with masses lower than 0.1$M_{Jup}$ is shown in the $top$ panel of Fig.~\ref{fig_p_m_low}.
As one can see, for a fixed planetary mass, most of the planets around metal-poor stars are constrained to longer periods.
This tendency appears to be valid from about 0.03$M_{Jup}$ to about 4$M_{Jup}$ and confirms and extends the 
results of \cite{Beauge-13} to higher masses.  

To evaluate the statistical significance of the observed trend we performed a simple Monte Carlo (MC) test.
First, we counted the number of planets orbiting metal-poor stars in a given interval of mass from 0.03$M_{Jup}$ to $M_{P}$(max), and we 
randomly drew the same number of points from the planetary sample around metal-rich stars.
Then we counted the number of planets orbiting metal-poor and metal-rich stars below the line of the linear fitting of the full data (P, M$_{P}$ of
all the planets in our initial sample, i.e., metal-poor and metal-rich planets with masses from 0.0036 to 17.4$M_{Jup}$).
We repeated the entire process 10$^5$ times. By comparing the average number (and using the standard deviation) of metal-rich planets found below 
the fitted line with the number of metal-poor planets (again bellow the fitted line), we evaluated the significance (the z-score - n$\sigma$) 
of the statement
that the two metal-poor and metal-rich populations have different distributions of (P,M$_{P}$)\footnote{
We note that the quoted significance levels are the result of a one-sided test, since the test by construction is one-sided.}%
. 

The $M_{P}$(max) originally was 1$M_{Jup}$, which we later varied up to the highest value of the planet mass in the sample. 
The metallicity boundary was also considered as variable (-0.10, -0.05, and 0.00 dex).
We found that the significance of the correlation decreases with the increase of $M_{P}$(max) and 
interestingly, after $M_{P}$(max) $=$ 4$M_{Jup}$, the significance decreased faster.
From the $right$ panel of Fig.~\ref{fig_p_m_feh} one can see that almost all the planets with masses higher than 4$M_{Jup}$ orbiting metal-poor stars
lie below  the dashed line. Depending on the combination of the parameters ([Fe/H] and $M_{P}$(max) $\leq$ 4$M_{Jup}$), the 
significance of our results varies from \{4.4$\sigma$ to 6.1$\sigma$, which means that it is always significant.

To ensure that the observed trend is genuinely valid for the higher-mass domain and is not just influenced by the previously reported trend for low-mass 
planets \citep{Beauge-13},
we repeated the test, selecting $M_{P}$(min) in the range from 0.1 to 0.3 Jupiter mass and excluding the planets with M$_{P}$ $>$ 4$M_{Jup}$. 
We confirm the previous result at a 1.8$\sigma$ to 3.7$\sigma$  significance level (depending
on the combination of fixed parameters, i.e., [Fe/H], $M_{P}$(min)): the significance was lowest when $M_{P}$(min) is 0.3 Jupiter mass and
the $M_{P}$(max) is 1$M_{Jup}$ (when the number of planets in the subsample is the smallest).
 
Furthermore we made a similar test, but instead of counting the planets below  the fitted line, we considered a horizontal line (constant period).
Varying this horizontal line from about 80 to 400 days, we again found that the result is always higher than 2.1$\sigma$ (reaching up to 3.2$\sigma$).
These tests show that the results do not depend significantly on the position of the separation (fitted line or constant period), and they suggest that the 
findings by \cite{Beauge-13} about the lack of small-size planets around metal-poor stars at short periods extend up to 4 Jupiter masses.

\subsection{Earth-like planets}

\begin{figure}
\begin{center}
\includegraphics[angle=0,width=1\linewidth]{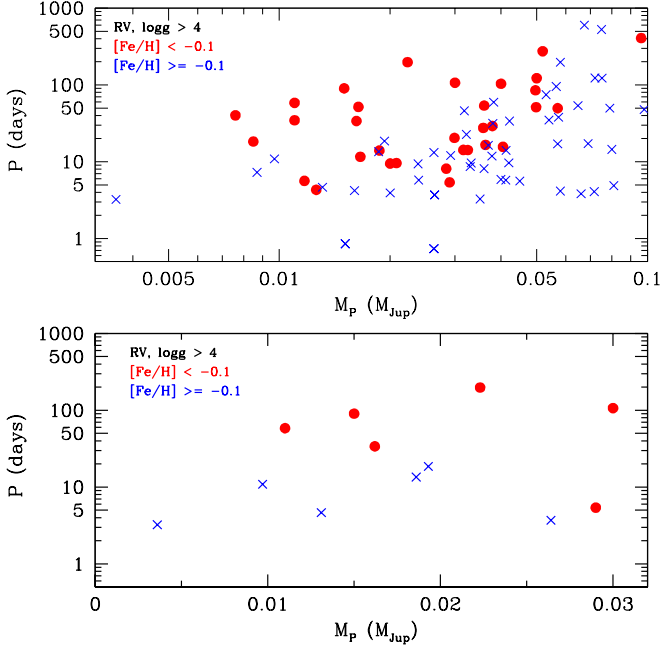}
\end{center}
\caption{Same as Fig.~\ref{fig_p_m_feh} right panel, but only for planets with M$_{P}$ $<$ 0.1$M_{Jup}$ ($top-panel$). The (P, M$_{P}$) diagram for the
longest-period planet in the system with M$_{P} <$ 0.03M$_{Jup}$ ($bottom-panel$).}
\label{fig_p_m_low}
\end{figure}

All the planets with masses bellow 0.03$M_{Jup}$ (about 10$M_{\oplus}$) orbiting metal-rich stars have short periods of fewer than 18 days 
(see $top$ panel of Fig. ~\ref{fig_p_m_low}). It is difficult to understand why these low-mass planets do not have longer periods, such 
as we observe for planets in the metal-poor systems. 
One of the reasons for this disposition can have a dynamical character. Most of the planets orbiting metal-poor stars belong to the multiple 
systems where almost all the planets have low masses (M$_{P} <$ 0.03$M_{Jup}$), while planets in the multiple systems orbiting metal-rich stars have
higher-mass planetary companions with longer periods. However, in the $bottom$ panel of Fig. ~\ref{fig_p_m_low}, where we show the distribution of the planets 
with the longest period in the system, one can see that there are six planets orbiting metal-rich stars without a (detected) higher-mass longer-period companion.
This means that at least for the planets presented in the plot, the shorter periods are not a result of interaction with higher-mass longer-period companions.
Another explanation of the lack of metal-rich low-mass planets with long periods might be a detection limit, since metal-rich planet hosts are on average slightly 
more massive and hotter than their metal-poor counterparts. However, the mentioned differences are very small and probably cannot be responsible for the 
observed short periods.

If there is no detection bias in the sample, then the observed distributions of metal-poor and metal-rich planets in the P-M$_{P}$ diagram could
mean that Earth-like planets orbiting metal-rich stars preferably migrate or form close to their parent stars, while planets in the metal-poor systems 
form at a wider range of the semi-major axis or do not always migrate. 
 
In the small range of periods (fewer than 18 days) where metal-rich planets orbit,  there are also similar planets orbiting metal-poor stars, 
and their period distribution seems similar (excluding the planets with longer periods orbiting metal-poor stars), although it is difficult
to evaluate the significance because of the small number of planets.

We note that the number of planets discussed in this subsection is small, therefore, the results and conclusions regarding them should be considered
with caution.

\subsection{[Ref/H] vs. [Fe/H]}

Typically, [Fe/H] is used as a proxy of overall metallicity for stars, but if a star has peculiar chemistry or is enhanced by some individual elements
compared to iron, then the [Fe/H] index will differ from the total metallicity [M/H]. \cite{Adibekyan-12a,Adibekyan-12b} showed that most of the 
planet-host stars with low-iron content are enhanced by $\alpha$-elements, including Mg and Si which are fairly abundant and have condensation
temperatures similar to iron \citep{Lodders-03,Lodders-09}. \cite{Beauge-13} recently questioned how their result would be affected 
if one used the [Ref/H] index\footnote{This index was proposed by \cite{Gonzalez-09} and quantifies the mass abundances of refractory elements (Mg, Si and Fe) 
relevant for planet formation.}
instead of [Fe/H].  

To quantify the effect of [Ref/H] on our result we used the HARPS sample of 135 planet-host stars with detailed and precise chemical abundances derived in
\cite{Adibekyan-12c}. In this sample 32 planet hosts have [Fe/H] lower than -0.1 dex. We found that in the mentioned metallicity region 
the [Ref/H] is on average higher than [Fe/H] by $\approx$ 0.05 dex: about 0.00 dex for stars with [Fe/H] $\approx$ -0.1 dex and up to 0.15 dex for the stars with
[Fe/H] $\approx$ -0.6 dex. 
Only two out of the mentioned 32 planet hosts with [Fe/H] $<$ -0.1 dex show [Ref/H] $>$ -0.1 dex, but both cases are very close to the boundary
(-0.09 and -0.08 dex).
Recalling that our results do not depend on the metallicity boundary, and assuming that the HARPS planet host sample is representative for the whole sample 
discussed above,  we can conclude that using the [Ref/H] index instead of [Fe/H] will not affect our results.

\subsection{Role of [Fe/H] on the position of planets in the P-M$_{P}$ diagram. Discussion}

The most important insights from our tests and results come from the continuity or discontinuity of the extrasolar 
planet population as a function of key parameters. The correlation between period, mass, and metallicity presented in 
Fig.~\ref{fig_p_m_feh} suggests the existence of a mechanism that affects a wide range of masses, from 10$M_{\oplus}$  to 4$M_{Jup}$, 
and that depends on metallicity. \cite{Dawson-13} showed that high-eccentricity Jupiter-mass planets with semi-major axes ($a$) between 0.1 and 1 AU were 
preferably found orbiting metal-rich stars; the authors interpreted this finding as evidence for planet-planet 
scattering on a population created by smooth migration. 
The main assumption was that [Fe/H] will not significantly affect the 
P-M$_{P}$ relationship as created by core-accretion+migration \citep[see][for details]{Mordasini-12}, and 
thus the observed difference was created exclusively by dynamical interactions.
They also noted that beyond 1 AU, the metal-rich and metal-poor samples have a similar eccentricity distribution with the 
explanation that planets with $a >$ 1AU may have formed where we observe them.

We applied a Kolmogorov-Smirnov (K-S) test for the samples of giant planets (M$_{P}$ $>$ 0.1$M_{Jup}$ and with P $>$ 10 days) orbiting metal-poor ([Fe/H] $<$ -0.1 dex)
and metal-rich ([Fe/H] $\geq$ -0.1 dex) dwarf stars and found that their eccentricity distributions are similar (K-S probability $\approx$ 0.63).
To test whether these distributions are similar for different period regimes we made two subsamples of planets 
with periods 10 $<$ P $\leq$ 300 days (300 days  corresponds to $\sim$1 AU, which was the upper limit of $a$ in \cite{Dawson-13}), 
and with P $>$ 300 days. Since the number of planets orbiting around
metal-poor stars with [Fe/H] $\geq$ -0.1 dex is small (only seven stars), we changed the metallicity boundary from -0.1 to 0.0 dex. We note that this was also
the metallicity boundary chosen by \cite{Dawson-13}. The applied K-S test 
predicts $P_{KS}$ = 0.10 and $P_{KS}$ = 0.97 that metal-poor and metal-rich populations of planets with short and long periods have the same eccentricity
distribution, respectively. Thus, we confirm the results of \cite{Dawson-13} for ''close-in`` giants. We note that for the whole range of periods (P $>$ 10 days) 
the K-S statistics delivers $P_{KS} \approx$ 0.66 for the similarity of eccentricity distribution of planets orbiting stars with [Fe/H] $<$ 0.0 dex and 
[Fe/H] $\geq$ 0.0 dex.
The test implies that the P-M$_{P}$ distribution of the metal-poor population (at least for planets with periods longer than 300 days)
results from formation and migration and has no dynamical character.
The fact that massive planets at low-[Fe/H] are found farther out implies that these planets probably are only able to form beyond the 
ice-line for low-metallicities \citep{Mordasini-10, Mordasini-12}\footnote{This would in turn imply that migration does not significantly change 
the semi-major axis of low-metallicity planets. This last point seems to be in line with what is expected from the models, 
but depends on the definition of the ice-line, among other aspects.}. 

The similarity of the period distribution of short-period Earth-like planets (M$_{P} < 10M_{\oplus}$) around metal-rich and metal-poor stars probably means that
enough low-mass planets can be formed at small semi-major axes even at low metallicities because a sufficient amount of protoplanetary mass is locally available.
At the same time, the fact that planets orbiting metal-rich stars are observed only at short periods might imply that these planets form close to their central stars
or ordinarily migrate.
However, we note that the statistical significance of the last results are not evaluated because of small number of planets.

It is also interesting to note that if we compare metal-poor stars with their metal-rich counterparts, there is a 
dearth of planets with mass around 0.1$M_{Jup}$ orbiting the metal-poor stars \citep[see also][]{Beauge-13}.
In the paradigm of \cite{Dawson-13}, these Neptune-mass planets 
are preferentially formed farther out, (again probably around or beyond the ice-line), and move into smaller orbits 
because of planet interaction. Interestingly, this is qualitatively in line with results from microlensing surveys, 
which state that cold Neptunes are indeed common \citep{Sumi-10, Gaudi-12}.
However, a quantitative analysis would require an assessment and correction for detection bias, which is far beyond the scope of this paper.


\begin{figure}
\begin{center}
\includegraphics[angle=270,width=1\linewidth]{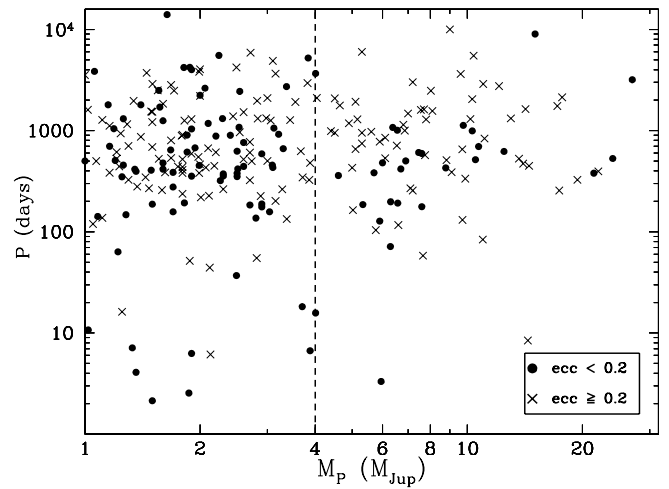}
\end{center}
\caption{(P,M$_{P}$) diagram for the all RV detected planets with M$_{P}$ $>$ 1$M_{Jup}$ orbiting FGK stars from \texttt{exoplanet.eu}. 
Planets with high- and low-eccentricity orbits are presented by crosses and circles, respectively.}
\label{fig_p_m_ecc}
\end{figure}

\section{Orbital and physical properties of very massive planets}

In the previous section we noted that the planets with very high masses show different dependencies on metallicity in the (P,M$_{P}$) diagram from
their less massive counterparts. This may be related both to formation and to post-formation processes (interaction with disk or interaction between planets).
To form a very high planetary mass, a critical core must grow, a process that takes longer in a low-metallicity environment \citep{Mordasini-12}.
The time scale of the orbital evolution of planets also depends on the properties of the protoplanetary disk and on other initial conditions
\citep[e.g.][]{Xiang-Gruess-13, Bitsch-13}. To determine whether this difference between high- and low-mass planets in the (P,M$_{P}$,[Fe/H]) diagram is reflected 
on other parameters as well, we compared the physical and orbital parameters of planets  with  masses from 1 to 4$M_{Jup}$ to those with masses 
higher than 4$M_{Jup}$\footnote{
We chose to establish a low-mass cut-off because one can assume that at least the mechanism of formation of the planets in this two planetary mass regimes 
is similar.}. For this analysis we used all the RV-detected planets orbiting FGK stars from the Extrasolar Planets Encyclopaedia \citep{Schneider-11}\footnote{
http://www.exoplanet.eu}.

\subsection{Orbital eccentricity}

The first significant correlation we found is that very massive planets are more eccentric than their lower-mass counterparts.
This correlation is illustrated in Fig.~\ref{fig_p_m_ecc}. Sixty-one out of 90 planets (66.3$\pm$5.0\%) with M$_{P}$ $>$ 4$M_{Jup}$ have $e$ $>$ 0.2, while only 94 out of 173
(54.3$\pm$3.8\%) planets with masses between 1 and 4$M_{Jup}$ are similarly eccentric. To assess the statistical significance of the difference 
in observed eccentricity distributions between low- and high mass planets we performed a K-S test. The K-S  statistics predict a 0.0026 ($\approx$3$\sigma$) 
probability ($P_{KS}$) that the two subsamples come from the same underlying distribution of $e$.  We note that when we consider only the planets orbiting FGK dwarf stars
the difference remains significant; 80.4$\pm$5.8 and 62.1$\pm$4.6 percent of eccentric planets with masses higher than 4$M_{Jup}$ and with 
1$M_{Jup}$ $<$ M$_{P}$ $<$ 4$M_{Jup}$, respectively. To test whether our result is affected by close-in planets, which mostly have circular 
orbits due to tide effects \citep[see e.g.][]{Ford-06}, we established a minimum period cut-off at 10, 50, and 100 days (although from Fig.~\ref{fig_p_m_ecc} it is
clear that most of the RV-detected planets have P $>$ 100 days). The K-S probabilities for the three cases 
were 0.004, 0.0036, and 0.0022 respectively, which allows us to conclude firmly that this is not the case.

\cite{Udry-02} and \cite{Ribas-07} had already found a marginal tendency for low-mass planets (M$_{P}$ $<$ 4$M_{Jup}$) to be less eccentric than more massive 
planets and binary stars. 
We note that in these two studies the authors did not separate very small planets (Earth-like or Neptune-like) from the massive gaseous planets\footnote{
It is worth to note, that if we compare eccentricities of low- and high-mass planets considering all the planets with M $<$ 4$M_{Jup}$ and M $>$ 4$M_{Jup}$
then the $P_{KS}$ decreases to 3$\times10^{-5}$.}.
However, \cite{Udry-02} noted that when one restricts the sample to periods longer than 50 days (avoiding the circularization through tidal interactions), the 
difference in eccentricities between low- and high-mass planets disappears.

It has been suggested that the eccentricity of planets can be increased through the planet-disk interaction under favorable conditions 
and especially if the planetary mass is very high \citep{Papaloizou-01, Kley-06, D'Angelo-06}. Very recently, \cite{Bitsch-13}, performing isothermal 3D simulations,
showed that the eccentricity of planets in single systems with masses between 1 and 5$M_{Jup}$  is generally damped due to planet-disk interaction, while 
for very massive planets with masses above $\sim$ 5$M_{Jup}$ the eccentricity can increase for low orbital inclinations relative to the disk.
At the same time, N-body simulations of multiple giant planets performed by \cite{Raymond-10} and population synthesis models by \cite{Ida-13}, and before that
by \cite{Thommes-08}, showed that
the eccentricity increases with planetary mass. The latter authors explained this with a scenario in which multiple giant planets are mainly formed in relatively
massive disks where dynamical instabilities, cohesive collisions, and orbit crossings are more common and can result in excitation of higher eccentricities.

Summarizing, \cite{Bitsch-13} predicted higher eccentricities for very high mass planets due to the interaction with disk (note that they performed 
their simulation for single-planet systems), and \cite{Ida-13} predicted correlation between $e$ and M$_{P}$ for a wider range of planetary mass because of
close scatterings and interaction of gas giants. We again applied a K-S statistics to test whether the eccentricity distribution of planets with
masses between 1 to 5$M_{Jup}$ and planets with M$_{P} >$ 5$M_{Jup}$ are similar or not, but now we considered only planets in single systems.
Our test predicts $P_{KS} \approx$ 0.004 that the two populations have the same underlying eccentricity distribution. We note that in the last test
we considered only planets with P $>$ 10 days. If one consider planets with P $>$ 50 or P $>$ 100 days, or changes the M$_{P}$ boundary to 4$M_{Jup}$, the 
$P_{KS}$ remains always smaller than 0.005 throughout.

Based on the observed trends, we can conclude that in general the eccentricities of high-mass planets, even in single-planet systems (although there is
a possibility that in the systems there are more planets that are undetectable by current instrumentation and surveys or that some planets have been ejected), are 
higher than those for lower-mass planets, as predicted from the numerical simulation \citep{Papaloizou-01, Bitsch-13}. 

Very recently, high-resolution near-infrared observation of HD100546 showed a variable ro-vibration of CO and OH emission lines 
\citep{Liskowsky-12, Brittain-13}, which the authors explained by postulating the presence of an eccentric massive planetary companion. This is an direct
observational support of our results obtained statistically for a large sample.

\subsection{Orbital period}

The second correlation we observe is that very massive planets with eccentric orbits have longer periods than those with more circular orbits with $e <$ 0.2 
(see Fig.~\ref{fig_p_m_ecc}).
The K-S  statistics predict 0.6\% probability that the two families of planets have the same period distribution. 
This difference in periods is probably related to the interaction and migration processes in the disk that the planets suffered.

\cite{Cumming-04} suggested that planets with long periods are on average easier to detect if their orbits are eccentric. However, our result cannot
be explained with this observational selection effect because there is no similar correlation observed for lower-mass planets. 
The K-S statistics do not report a significant difference in period for planets with eccentric and circular orbits and 
with masses 1$M_{Jup}$ $<$ M$_{P}$ $<$ 4$M_{Jup}$. 
The $P_{KS}$ is 0.15 that low- and high-eccentricity planets with a mass in this interval have the same period distribution . 
This probability even increases from 0.15 to 0.76 when restricting the sample to periods longer than 10 to 100 days.
Since the period distribution of planets with low (1$M_{Jup}$ $<$ M$_{P}$ $<$ 4$M_{Jup}$) and high mass is similar ($P_{KS} \approx 0.65$ for planets with 
P $> 10$ days), one could expect a more prominent correlation between eccentricity and period for lower-mass planets if there is a selection bias. 
This allows us to conclude that the correlation between $e$ and P for very high mass planets has a physical meaning, and does not come
from an observational bias.

\subsection{Metallicity}

Interestingly, \cite{Ribas-07} found that the metallicity of planet host stars decreases with planet mass. In particular, they showed that the average metallicity
of stars with planets of mass lower than 4$M_{Jup}$ is different from that of higher mass planet hosts at 3-$\sigma$ level, which is lower by about 0.15 dex. 
We again note that in their comparison
they did not separate super-Earths and Neptunes, and in their sample almost all of the very low mass planets (M$_{P}$ $<$ 0.3$M_{Jup}$) have [Fe/H] $>$ 0, 
while the majority of recently detected low-mass planets have subsolar or solar metallicities \citep[e.g.][]{Mayor-11, Buchhave-12}.

\begin{figure}
\begin{center}
\includegraphics[angle=0,width=1\linewidth]{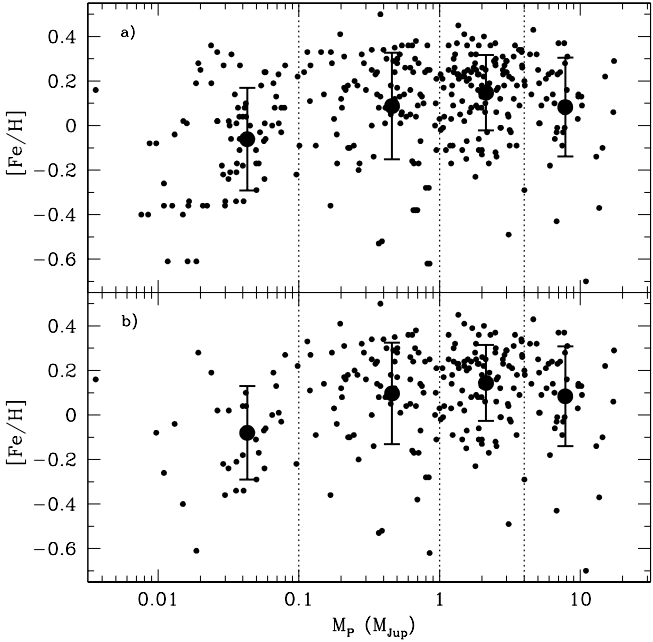}
\end{center}
\caption{Planetary mass against metallicty a) for all the planets in the system and b) for the most massive  planets in each planetary systems.
All the planets are detected with the radial velocity technique and planet hosts are FGK dwarf stars.
The small circles correspond to individual planets and the large circles represent the average metallicty for the four mass bins
M$_{P} < 0.1M_{Jup}$; $0.1M_{Jup} \leq M_{P} < 1M_{Jup}$; $1M_{Jup} \leq M_{P} < 4M_{Jup}$; and M$_{P}$ $>$ 4$M_{Jup}$. The error bars indicate
the standard deviation of the [Fe/H] in each bin.}
\label{fig_mass_feh}
\end{figure}

Figure~\ref{fig_mass_feh} shows the metallicities of FGK dwarf stars hosting RV-detected planets as a function of the planet mass. The metallicities were 
taken from the SWEET-Cat. On average, the metallicities of high-mass planets are similar. The average values and the standard error
of the means are presented in the Table 1. 

We tested the results obtained by \cite{Ribas-07} by applying a K-S test for the RV-detected sample of planets with masses between 1 and 4$M_{Jup}$ and 
with M$_{P}$ $>$ 4$M_{Jup}$. We obtained $P_{KS}$ $\approx$ 5.5\% that the two subsamples have the same underlying metallicity distribution. 
If we consider only the planet with highest mass in the system, the probability increases to 15.3\%. We note that 
the metallicities of these stars were taken from the literature, are not homogeneous, and include stars with different evolutionary stages.
We again applied  the same statistics, but taking only RV-detected planet-host FGK dwarf stars for which SWEET-Cat provides homogeneous stellar parameters, 
including metallicity. In this case the difference in metallicity distribution between the two subsamples becomes even less significant with  $P_{KS}$ $\approx$ 15\%
and $\approx$ 34.5\% if one consider only the planet with the highest mass in each system.
We therefore conclude that a simple separation in mass at 4$M_{Jup}$ does not reveal two different populations in metallicity, 
as hypothesized in \cite{Ribas-07}. This result is expected from the core-accretion theory, since the 
metallicity mostly acts as a threshold for giant planet formation, but is not correlated with the mass of giant planets, except for planets with 
M$_{P}$ $\gtrsim$ 10-20$M_{Jup}$, which are harder to form at clearly subsolar metallicities \citep{Mordasini-12}.  

\begin{table}
\centering
\caption{Average and standard errors of [Fe/H] for planets with different masses: taking into account all the planets in the system (a) and 
only the most massive planet in the systems (b).}
\label{ks}
\begin{tabular}{lcc}
\hline
\hline 
Planet mass & $<$[Fe/H]$>_a$ & $<$[Fe/H]$>_b$\tabularnewline
$M_{Jup}$ & dex & dex\tabularnewline
\hline 
M$_{P} < 0.1$  & -0.061$\pm$0.025 & -0.082$\pm$0.036\tabularnewline
$0.1 \leq M_{P} < 1$  & 0.088$\pm$0.023 & 0.096$\pm$0.025\tabularnewline
$1 \leq M_{P} < 4$  & 0.149$\pm$0.016 & 0.146$\pm$0.018\tabularnewline
M$_{P}$ $>$ 4  & 0.083$\pm$0.032 & 0.084$\pm$0.033\tabularnewline
\hline 
\end{tabular}
\end{table}




\section{Summary}

We analyzed the possible relations between orbital and physical properties of planets with different mass and physical
properties of their host stars. The main findings and results are itemized below .
\begin{itemize}

\item 
We found that for a fixed maximum planetary mass between 10M$_{\oplus}$ and 4$M_{Jup}$, planets orbiting metal-poor stars are constrained 
to longer periods than those orbiting metal-rich stars. Applied MC tests show that the obtained correlation is statistically significant. 
This result suggests that the mechanism responsible for the ''separation'' of planets in the P-M$_{P}$ is operational for a wide range of planetary mass. 
The observed dependence can be explained by assuming that planets in a metal-poor disk form farther out from their central stars and/or do not migrate 
as far as planets in metal-rich systems because they form later.
Our result confirms the \cite{Beauge-13} findings and extends their conclusions to higher planetary masses.

\item
The Earth-like planets (M$_{P} < 10M_{\oplus}$) orbiting metal-rich stars have shorter periods than those orbiting metal-poor stars.
If there is no detection bias, this could imply that the low-mass planets in metal-rich systems ordinarily migrate or that they always form close
to their parent stars. 
The presence of Earth-like planets with short-periods orbiting metal-poor stars probably means that these planets can be formed close to 
their parent stars even at low metallicities because enough amount of protoplanetary mass is locally available.

\item

By applying a K-S test for a sample of FGK RV-detected dwarf planet-hosts we obtained $P_{KS}$ up to 35\% that the low- and high-mass 
planet hosts have the same underlying metallicity distribution. This result is expected in the core-accretion models \citep[e.g.][]{Mordasini-12}, 
but it contradicts the results of \cite{Ribas-07}, who found that the metallicity of planet-host stars decrease with planet mass.

\item
Analyzing the eccentricity distribution of high-mass planets, we found statistically significant evidence
that planets with masses higher than 4$M_{Jup}$ have on average more highly eccentric orbits than the giant planets with masses between 1$M_{Jup}$ and 4$M_{Jup}$.
These trends and dependencies agree with the core-accretion and formation in a disk models and allow us to conclude that 
planet-disk interaction is a very important and orbit-shaping mechanism when one moves towards higher planetary masses.    
In addition to these observational results, we found that less eccentric very high mass planets (M$_{P}$ $>$ 4$M_{Jup}$) have shorter periods than those with 
similar mass that orbit their host stars with more highly eccentric orbits. 
This difference in periods is probably related to the interaction and migration processes in the disk that the planets underwent.

\end{itemize}

The dependencies and trends presented in this work can provide new constraints for the models and numerical simulations of planet formation and evolution.
In particular the fact that almost all the giant planets orbiting metal-poor stars show long periods (i.e. longer than 100 days) shows that
migration is less rapid than assumed in core-accretion planet formation models \citep[e.g.][]{Mordasini-12}.

%
\begin{acknowledgements}

{This work was supported by the European Research Council/European Community under the FP7 through Starting Grant agreement 
number 239953. V.Zh.A., S.G.S., E.D.M., and M.O. are supported by grants SFRH/BPD/70574/2010, 
SFRH/BPD/47611/2008, SFRH/BPD/76606/2011, and SFRH/BD/51981/2012 from the FCT (Portugal), respectively. 
A.C. is supported by grant PEst-C/CTM/LA0025/2011 from the FCT.
C.M. acknowledges the support of the MPG through the Reimar-L\"ust Fellowship.
G.I. acknowledges financial support from the Spanish Ministry project MINECO AYA2011-29060.
We gratefully acknowledge the anonymous referee for the constructive comments and suggestions, and Astrid Peter for
the help concerning English.}
\end{acknowledgements}

\bibliography{refbib}

\end{document}